\documentclass[showpacs,floatfix,superscriptaddress,showpacs,preprint,amssymb,amsfonts,prb,aps]{revtex4}
\usepackage{longtable,graphicx,epsfig,dcolumn}

\begin{document}
\bibliographystyle{revtex}
\title{Surface Phases in Binary Liquid Metal Alloys: 
An X-ray Study }
\author{Holger~Tostmann}
\affiliation{Division of Applied Sciences and Department of Physics,
Harvard University, Cambridge MA 02138}

\author{Elaine~DiMasi}
\affiliation{Department of Physics, Brookhaven National Laboratory,
Upton NY 11973-5000}

\author{Oleg~G.~Shpyrko}
\affiliation{Division of Applied Sciences and Department of Physics,
Harvard University, Cambridge MA 02138}

\author{Peter~S.~Pershan}
\affiliation{Division of Applied Sciences and Department of Physics,
Harvard University, Cambridge MA 02138}

\author{Benjamin~M.~Ocko}
\affiliation{Department of Physics, Brookhaven National Laboratory,
Upton NY 11973-5000}

\author{Moshe~Deutsch}
\affiliation{Department of Physics, Bar-Ilan University, Ramat-Gan
52100, Israel}

\begin{abstract}

Surface sensitive x-ray scattering techniques with atomic scale
resolution are employed to investigate the microscopic structure of
the surface of three classes of liquid binary alloys: (i) Surface
segregation in partly miscible binary alloys as predicted by the
Gibbs adsorption rule is investigated for Ga-In. The first layer
consists of a supercooled In monolayer and the bulk composition is
reached after about two atomic diameters. (ii) The Ga-Bi system
displays a wetting transition at a characteristic temperature $T_w
\approx$ 220$\rm^o$C. The transition from a Bi monolayer on Ga below
$T_w$ to a thick Bi-rich wetting film above $T_w$ is studied. (iii)
The effect of attractive interactions between the two components of
a binary alloy on the surface structure is investigated for two
Hg-Au alloys.

\end{abstract}


\maketitle

\section{Introduction}

Theoretical calculations and computer simulations indicate that the
microscopic structure of the surface of liquid metals (LM) and
alloys is considerably different from the surface structure of
dielectric liquids and mixtures\,[1-3]. For liquid metals, the
density profile normal to the surface has been predicted
theoretically to show  oscillations with a period of about one
atomic diameter and extending a few atomic diameters into the
bulk\,[2]. This layering normal to the surface has been confirmed
experimentally for liquid Hg [4], Ga\,[5] and, most recently,
In\,[6]. Surface induced layering in LM is is due to the drastic
changes in the interactions across the interface from short-ranged
screened Coulomb interactions in the liquid phase to long-ranged van
der Waals interactions in the gas phase. In dielectric liquids, on
the other hand, van der Waals interactions prevail both in the
liquid and in the gas phase and theory predicts a  monotonic density
profile\,[7].

Compared to elemental LM, binary LM alloys have an additional
degree of freedom and it is possible  to study the effect of the
second component on surface induced layering.
Moreover, binary mixtures are expected to
exhibit a richer array of surface structures.
In  partly miscible mixtures for example, the lower surface tension
component is expected to segregate at the surface as predicted
by the Gibbs adsorption rule of thermodynamics\,[8].
 For binary liquid metal alloys,
computer simulations predict that
the component that segregates
at the liquid-vapor interface forms a nearly pure monolayer\,[9-11]. The
adjacent layer on the bulk liquid side is slightly depleted of the
surface-active component, and the bulk composition is reached at
a distance of about two atomic diameters into the bulk.
For simple dielectric mixtures on the other hand, theory predicts
that the excess concentration of the segregating component falls
smoothly with distance into the bulk over a range of several atomic
diameters\,[8].
These predictions remain largely untested experimentally.

The bulk phase behavior of  demixing leading to a miscibility gap with
a critical consolute point
induces a new class of
surface phase transitions as predicted by Cahn in 1977\,[12].
Based on scaling law arguments Cahn postulated
that a phase transition from nonwetting of the two immiscible phases
to complete wetting of one phase by the other necessarily occurs
close to the critical consolute point.
In this transition, a macroscopically thick wetting phase is formed
intruding between the bulk liquid and the vapor phase (or
an inert substrate). This is in contrast to the phenomenon of surface segregation
that takes place on atomic length scales.
Experimental studies of such wetting phase transitions have been restricted
mostly to
binary dielectric mixtures\,[13].
Wetting and prewetting phase transitions have been observed only recently in
fluid systems governed by screened Coulomb interactions\,[14], namely
K-KCl\,[15], Ga-Pb\,[16] and  Ga-Bi\,[17].
These experiments provide no direct microscopic information on the structure of the interface.
By contrast, the study presented here probes the structure of the liquid
interface with
{\rm \AA} resolution using surface
x-ray scattering techniques.

A  characteristic feature of many well-known and technically important alloys is
the  formation of intermetallic phases in the solid state\,[18].
These phases include stoichometric Laves
phases, broad Hume-Rothery phases and semiconducting Zintl phases
containing polyanions.
This phase formation is due to attractive interactions between the two
components. An interesting question is whether these attractive interactions
affect the surface structure of liquid binary alloys.

In this paper, we will demonstrate how surface x-ray scattering techniques
can be used to study LM alloys exhibiting different types of surface structure.
In Sections 2 and 3 we review the x-ray scattering techniques and
experimental details. In Section 4 we present three different alloy
systems with increasing complexity of bulk phase
behavior and  surface structure: Ga-In, Ga-Bi
and Hg-Au.

\section{Surface X-ray Scattering Techniques}

X-rays are well suited to study the structure of the surface of liquids
with atomic scale resolution.
The large dynamic range in intensity required for such studies can be
realized to advantage by synchrotron x-ray sources. In this section
we describe the surface x-ray scattering techniques used in our studies
to determine the structure normal to the surface and within the surface
plane.

\subsection{ 2.1. Specular X-ray Reflectivity (XR):} X-ray
reflectivity probes the  structure normal to the  interface. Here,
the x-ray intensity is measured at the specular condition where
incoming and outgoing angles $\alpha$ and $\beta$ are equal and the
azimuthal angle $2\theta$ is zero (see Fig.\,1). In this case, the
scattering momentum transfer, $\vec{q} =
\vec{k}_{out}-\vec{k}_{in}$, is normal to the surface
\begin{equation}
|\vec{q}| = q_z = \frac{2\pi}{\lambda} (\sin \alpha +  \sin \beta) = \frac{4\pi}{\lambda} \sin
\alpha
\end{equation}
where $\lambda$ is the x-ray wavelength.

If the liquid surface is ideally flat and abruptly terminated (i.e. has
a step function density profile), the reflectivity $R$  is given by the Fresnel
reflectivity $R_f$ known from classical optics\,[19].
The reflectivity of a real surface will deviate from $R_f$. For example, height
fluctuations produced by thermally excited capillary waves cause phase
shifts in x-rays reflected from different points on the surface.
Averaging over these  phase variations  gives rise to a
Debye-Waller type factor, $\exp(-\sigma_{cw}^2 q_z^2)$,
characterized by a capillary wave roughness $\sigma_{cw}$\,[6].
In addition, the density profile normal to the surface deviates from a
step function, being oscillatory for elemental LM.
For binary alloys displaying phase separation or phase transitions
we expect  further modification of the surface. Therefore,
a surface structure factor $\Phi$ has to be taken into account\,[20]:
\begin{equation}
\label{master}
\Phi (q_z) =  \frac{1}{\rho_{\infty}} \int_{-\infty}^{+\infty}
\frac{d \langle \rho (z) \rangle }{d z} \exp(\imath q_z z)dz
\end{equation}
with $\langle \rho (z) \rangle$ denoting the electron density average over a microscopic
surface area on the
atomic scale and $\rho_{\infty}$ the bulk electron density.
This factor is analogous to the bulk structure factor $S(\vec{q})$ which is
the Fourier transform of the bulk electron density. The reflectivity from
a real surface is then given by:
\begin{equation}
\label{final}
R(q_z) = R_f \exp(-\sigma_{cw}^2 q_z^2) \left| \Phi (q_z)\right|^2
.\end{equation}
It is not possible to directly invert the measured reflectivity $R(q_z)$ to the
density profile normal to the surface due to the loss of phase information.
Rather, the common practice is the construction of a physically plausible model
for the density profile which is then inserted in eq.\,(\ref{master}) and fitted
to the measured reflectivity\,[6,21].

\subsection{Grazing Incidence Diffraction (GID):}

GID probes the in-plane structure of the surface.
The in-plane momentum transfer $q_{xy}$ is probed
by varying the azimuthal
angle $2\theta$ (see Fig.\,1).
This geometry is surface sensitive if the incident angle $\alpha$
is kept below the critical angle for total
external reflection $\alpha_{crit}$ thereby limiting the
x-ray penetration depth to $\simeq 50${\AA} and minimizing the
background from bulk scattering\,[22,23].
The GID surface in-plane structure can be directly compared to the bulk
liquid structure by making similar measurements for $\alpha > \alpha_c$.

\section{Experimental}

It is essential  to establish a clean oxide free liquid alloy surface since
even microscopic impurities significantly change the x-ray reflectivity\,[24].
We use slightly different methods to prepare the three alloys presented in
this paper.
The eutectic Ga-In alloy
is prepared in a High Vacuum chamber using in situ glow-discharge to clean the
sample\,[21]. The sample is transferred into the x-ray UHV chamber, melted
and cleaned by Ar ion sputtering.

The Ga-Bi alloy is prepared as follows.
Bi is  melted under UHV conditions and any residual oxide is evaporated by heating the
Bi to about 600$^{\circ}$\,C. After the Bi is transferred into an Ar
 glove box, liquid Ga is added so that  the  mole fraction of
Bi is $x_{Bi}\approx 0.3$. This value is chosen
to correspond to the concentration of the bulk solution at the critical point of 265$^{\circ}$\,C .
The sample is then frozen and transferred to the x-ray UHV chamber. The chamber
is baked out and the liquid surface sputter cleaned.

The preparation of  the Hg-Au alloy is quite different since
the high vapor pressure of Hg is not compatible with UHV conditions.
Quadruple distilled Hg and Au powder are mixed in a glove box and
transferred to a stainless steel reservoir attached to a valve with a filling
capillary. This container is attached to a UHV chamber with the valve closed.
This chamber is then subjected to a standard UHV bakeout to remove oxygen
and water, after which the valve is opened and the Hg-Au alloy is poured
into the sample pan. Due to the low oxygen partial pressure in the chamber,
residual oxide is instable and disappears from the surface within a couple of hours.

The sample chamber is placed on an active vibration
isolation table that effectively quenches all mechanically induced vibrations\,[6].
The experiments have been performed at the beamlines X22B and X25 at the National
Synchrotron Light Source at Brookhaven National Laboratory. The x-ray energies
are 19keV and 10keV at X25 and X22B respectively. A discussion of the liquid surface
spectrometer can be found elsewhere\,[21].

\section{Results and Discussion}

\subsection{Surface Segregation in Ga-In}

As an example of a simple miscible alloy with a eutectic point we
have studied the  Ga-In (16.5\% In) system\,[21]. The
Fresnel-normalized  reflectivity
is shown in Fig.\,2 (filled squares). The data have been recorded
at room temperature and are compared to the normalized reflectivity
from elemental Ga (open squares) at the same temperature.
The most pronounced feature of the Ga reflectivity is the quasi-Bragg
peak at 2.4\,${\rm \AA}^{-1}$ that indicates the presence of surface induced
layering\,[5]. This layering peak is present in the Ga-In alloy as well,
although slightly suppressed. The peak position is centered at a smaller
value of $q_z$ compared
to Ga indicating a larger layer spacing, as expected for a larger In atom.
The second pronounced difference between the Ga-In and the Ga reflectivity
is the observation that the reflectivity from Ga-In is  about
30\% higher than the
Fresnel reflectivity of an ideal Ga-In surface at intermediate values
for $q_z$. This can only be understood if a higher density adlayer exists
at the surface.
The measured x-ray reflectivity is
consistent with the segregation of a 94\% In monolayer on top of the  Ga-In eutectic.
The presence of the In monolayer increases the reflected x-ray
intensity at intermediate $q_z$ due to the higher electron density of In but has a
less pronounced effect at larger $q_z$ values where the reflectivity is dominated
 by the Ga layering peak.
The density profile that fits the measured XR
can be modeled by a single Gaussian for the In segregation layer plus
a semi-infinite sum of Gaussians for the Ga-In eutectic with half widths that increase
with distance from the surface, representing the decay of the  layering
with increasing distance from the surface\,[21]. The best fit of
this model to the data results in a density profile normal to the surface that shows
the segregation of an In monolayer with a 2.6{\AA} spacing between the In monolayer
and the underlying eutectic layer. This fit is represented by the
solid line in Fig.\,2.

These results provide new information about the surface structure of miscible
LM alloys. The constituents of this alloy have similar size and the same valency.
The Ga-In interactions are comparable
to the Ga-Ga and the In-In interactions. As a consequence, Ga-In is close to being
an ideal mixture, for which surface segregation of the lower surface tension
component is predicted. Our measurements show that this segregation is actually confined
to a single monolayer.
An arguably more interesting issue is the interplay between surface segregation
and surface induced layering. Our measurements show that layering persists,
with the In monolayer comprising the outermost layer. Similar conclusions
were reached by recent Quantum Molecular Dynamics simulations\,[25].

\subsection{Wetting phase transition in Ga-Bi}

The Ga-Bi system  is an example of an alloy  with a miscibility gap.
Below the monotectic temperature, $T_{mono}$= 222$^{\circ}$\,C, a
Ga rich liquid coexists with a solid Bi phase.
(see\,[26] for a full description of the phase diagram).
However, due to its lower surface energy a Bi monolayer is
expected to segregate at the surface of the Ga rich liquid.
The  wetting phase transition that is predicted for all binary mixtures with
critical demixing\,[12]  occurs at a characteristic wetting temperature
$T_w$ below the critical temperature $T_{crit}$\,[27,28]. Optical studies show that
$T_w$ coincides with the monotectic
temperature  for Ga-Bi\,[14,17].
Above $T_w$, a macroscopically thick Bi rich phase is expected to completely
wet the lighter Ga rich phase in defiance of gravity.
The Bi concentration in the Ga rich phase increases with
increasing temperature as long as the liquid coexists with the solid Bi phase.

The normalized x-ray reflectivity spectra, $R/R_{f,Ga}$, for Ga-Bi at
35$^{\circ}$\,C and 95$^{\circ}$\,C  are shown in Fig.\,3 versus $q_z^2$.
At the lowest values of $q_z$, the normalized reflectivity approaches
the reflectivity of an ideal Ga surface given by the Fresnel
Law of Optics.
At values of $q_z$ close to the critical q-vector
$q_{crit}=4\pi/\lambda \sin \alpha_{crit}$, the reflectivity is not
particularly sensitive to the structure
or roughness of the surface (see eq.(\ref{final})) and is simply related to the
electron density within $\approx$20\,{\AA} of the surface.
The fact that the reflectivity approaches the Fresnel
reflectivity given by the Ga electron density implies that the higher density Bi can
not be more than a few monolayers thick.
 With increasing $q_z$,
the normalized reflectivity first increases and reaches a maximum at
about $q_z \approx 1 {\rm \AA}^{-1}$. The variation of $R/R_{f,Ga}$ with
$q_z$, is similar to that observed by Lei et al.\,[10],
corresponding to the thickness of a single atomic layer, again
suggesting that only the top layer has a sizeable bismuth concentration. Compared to the
Ga-In alloy discussed above, this variation is much more pronounced. This results from
the 30\% higher electron density of bismuth compared with  gallium, and only
a 5\% higher density for indium.  The behavior of the
same alloy at 260$^{\circ}$C is markedly different. Here, the reflectivity
starts out at $R/R_{f,Ga} \approx 2$ and falls off monotonically with $q_z$,
a behavior consistent with a thick Bi rich layer terminating the metal/vapor
interface above $T_w$.

The reflectivity below $T_w \approx 220^{\circ}$C has been analyzed
in terms of a density profile similar to the one which describes
the Ga-In data\,[21]. The best fits, the solid lines in Fig.\,3, provide an
excellent description of the data.  The corresponding local density profile,
shown in Fig.\,4 after removing the temperature dependent capillary
wave roughness factor, exhibits a top-layer density which is about 1.5 times
higher than the Ga bulk liquid density and is consistent with
the formation of a complete Bi monolayer. The 3.6$\pm 0.2 {\rm \AA}$ layer
spacing between the monolayer and the adjacent Ga layer (obtained from
the fits) is much larger than the $2.5\pm 0.1 {\rm \AA}$ layer spacing obtained
in liquid gallium.  This also supports the conclusion of a top layer with a much larger
atomic diameter.
In addition, the 4.3{\AA} exponential decay length of the layering amplitude
is significantly smaller than the 5.5{\AA} decay length obtained for pure liquid Ga\,[5].
The suppression of the layering in the presence of either a bismuth or indium
monolayer is much more apparent in the case of bismuth (compare Fig.\,3 to Fig.\,2).
A likely explanation is the larger size difference in Ga-Bi
(the Bi diameter is 30\% larger than that of Ga) compared to Ga-In (In is
15\% larger than Ga).
As demonstrated by the solid lines, the layering model which describes
the 35$^{\circ}$C reflectivity profile also agrees closely with
 the 95$^{\circ}$C profile, without a discernible change in the parameters of the density profile,
except for the value of $\sigma_{cw}$, which is scaled, using capillary wave theory, to
correspond to the higher temperature (see eq.(\ref{final})).

This picture of a Bi monolayer segregating on top of the Ga rich bulk phase
is supported by GID experiments which probe the in-plane structure of the
surface (see Fig.\,5). The solid line shows the in-plane liquid structure
factor measured for $\alpha > \alpha_{crit}$ where the x-rays penetrate the bulk to
a depth $>1000${\AA}. The broad peak at $q_{xy} \approx 2.5 {\rm \AA}^{-1}$
and the shoulder on the high-angle side of the peak  agree with
bulk liquid Ga structure factor\,[29].  There is no evidence for a peak or shoulder at the
position corresponding to the first peak of the Bi liquid structure factor at
$q_{xy} \approx 2.2 {\rm \AA}^{-1}$.
The bulk scattering completely  masks
the weak scattering from the bismuth monolayer due to the large penetration depth of x-rays.
The surface sensitivity is drastically enhanced by keeping the incoming
angle $\alpha$ below the critical angle for total external reflection
(0.14$^{\circ}$ for Ga). As the x-rays now penetrate only to a depth $<50$\,{\AA}, the first peak of
the Bi liquid structure
factor is clearly visible at $q_{xy} \approx 2.2 {\rm \AA}^{-1}$ (open triangles in Fig.\,5).
Still, since the Bi layer is only a single monolayer thick, the contribution from the
underlying Ga  is larger then that of the Bi, as seen from the Bi being only a shoulder on
the Ga structure factor peak in Fig. 5.

While the XR results at 35$^{\circ}$\,C and 95$^{\circ}$\,C are consistent with a real
space model where the Bi is  segregated only in the top atomic layer,
at temperatures above 220$^{\circ}$\,C, corresponding to a Bi mole fraction
of about 0.08, this model no longer describes the data.
The ratio $R/R_{f,Ga}$ for the reflectivity taken at 260$^{\circ}$\,C close
to the critical wavevector for Ga is approximately twice as large
as the same ratio taken at 35$^{\circ}$\,C and 95$^{\circ}$\,C.
The dashed horizontal line, shown in the inset to Fig.\,3, represents the
theoretical Fresnel-normalized reflectivity curve for bulk Bi ($R/R_{f,Bi} \approx 2\times
R/R_{f,Ga}$).  This curve  agrees well with  the 260$^{\circ}$ data at
small $q_z$ where the surface roughness contribution is minimal, and clearly supports the
conclusion
of a  thick Bi-rich surface layer. Such a thick wetting
layer has been proposed on the basis of
previous optical experiments\,[14,17].  These findings are also supported by GID
experiments above 220$^{\circ}$\,C which show a pronounced increase in the
intensity of the Bi structure factor relative to the Ga structure factor\,[30].

\subsection{Pair formation in Hg-Au}

Our x-ray reflectivity measurements of dilute liquid Hg-Au alloys
demonstrate that dramatic differences in surface structure can be
induced by very small changes in concentration.
Fig.~6(a) compares the reflectivity of Hg with that of a solution
of 0.06at\% Au in Hg.
This composition is approximately half of the reported solubility
limit for Au in Hg at room temperature [26].
For pure Hg, the surface layering peak at $q_z=2.2$~\AA\ is much
more prominent close to its melting point at $-40^\circ$C than at
room temperature
[31].
This difference arises primarily from increased roughening of the
surface due to thermally excited capillary waves with increasing temperatures.
The presence of 0.06at\% Au sharply suppresses the surface induced
layering of Hg, as shown by the substantial attenuation of the
layering peak.
The temperature dependence in both systems suggests that at this
concentration, the suppression of the surface layering by the Au
has a small dependence on temperature in comparison to the
effect of capillary waves.

Increasing the Au concentration to 0.13at\% yields  a qualitatively
different behavior as shown in  Fig.~6(b).
At room temperature, layering is suppressed to a similar extent in
both alloys.  However, upon lowering the temperature
of the 0.13at\% solution to $-30^\circ$C, a deep minimum appears
at $q_z=1.4$~\AA$^{-1}$.  This minimum is produced by destructive
interference between x-rays reflecting from layers of different density
near the surface, implying that one or more phases have formed that are
different from the composition of the bulk.
We find significant hysteresis of the reflectivity under temperature
cycling.  In particular, on returning to room temperature and cooling
again, the low temperature minimum is much less deep than that observed
previously, and saturates at this higher value even when the sample is
supercooled to about $-70^\circ$C.
This pronounced effect has been reproduced and in the absence of detailed
modeling we can only speculate about its origin.
As opposed to Ga-In and Ga-Bi, the
interactions between Hg and Au are attractive leading to intermetallic
phase formation in the solid state with $\rm Au_{2}Hg$ being the best
known phase\,[26]. Several other phases have been reported
that are only  stable below -39$^{\circ}$C\,[26].
It is known from x-ray studies on
 FeCo (001)\,[32] and $\rm Cu_3Au$ (001)\,[33] surfaces, that
this pair formation gives rise to mesoscopic surface segregation profiles in the solid state.
As soon as one component segregates at the surface, it is energetically
favorable for the next layer to consist entirely of the other component
due to the pronounced attractive interactions. It is conceivable that a
similar mechanism prevails in liquid Hg-Au. An alternating or more
complicated segregation profile would destroy the Hg layering even
more effectively than the segregation of a monolayer in Ga-In or
Ga-Bi and is consistent with our experimental findings that a small
amount of Au completely suppresses the otherwise very pronounced Hg
layering peak even at low temperatures.

\section{Acknowledgements} This work has been supported by the U.S.
Department of Energy Grant No. DE-FG02-88-ER45379,  the National
Science Foundation Grant No. DMR-94-00396 and the U.S.--Israel
Binational Science Foundation, Jerusalem. Brookhaven National
Laboratory is supported by U.S. DOE Contract No. DE-AC02-98CH10886.
HT acknowledges support from the Deutsche Forschungsgemeinschaft.

\section*{References}

[1] M.Iwamatsu and S.K.Lai, J.Phys.:Condens.Matter {\em 4}, 6039
(1992).\\

[2] J.G.Harris, J.Gryko and S.A.Rice, J.Chem.Phys. {\em 87}, 3069 (1987).\\

[3] R.Evans and M.Hasegawa, J.Phys.C:Solid\,State\,Phys. {\em 14}, 5225
(1981).\\

[4] O.M.Magnussen, B.M.Ocko, M.J.Regan, K.Penanen, P.S.Pershan
and M.Deutsch, Phys.Rev.Lett. {\em 74}, 4444 (1995).\\

[5] M.J.Regan, E.H.Kawamoto, S.Lee, P.S.Pershan, N.Maskil, M.Deutsch,
O.M.Magnussen, B.M.Ocko and L.E.Berman, Phys.Rev.Lett. {\em 75}, 2498 (1995).\\

[6] H.Tostmann, E.Dimasi, P.S.Pershan, B.M.Ocko, O.G.Shpyrko and
M.Deutsch, submitted (1998).\\

[7] G.A.Chapella et al., J.Soc.FaradayII {\em 73}, 1133 (1977).\\

[8] J.S.Rowlinson and B.Widom, Molecular Theory of Capillarity, Clarendon, Oxford, 1982.\\

[9] J.G.Harris and S.A.Rice, J.Chem.Phys. {\em 86}, 7531 (1987). \\

[10] N.Lei, Z.Huang and S.A.Rice, J.Chem.Phys.{\em 104}, 4802 (1996).\\

[11] S.A.Rice, M.Zhao and D.Chekmarev, submitted (1998).\\

[12] J.W.Cahn, J.Chem.Phys {\em 66}, 3667 (1977).\\

[13] M.Schick in: Liquids at Interfaces (Proceedings Les Houches Summer School, Session XL),
ed. by J.Charvolin, J.F.Joanny and J.Zinn-Justin, North-Holland, Amsterdam, 1990.\\

[14] W.Freyland and D.Nattland, Ber.Bunsenges.Phys.Chem. {\em 102}, 1 (1998).\\

[15] H.Tostmann, D.Nattland and W.Freyland, J.Chem.Phys. {\em 104}, 8777 (1996).\\

[16] D.Chatain and P.Wynblatt, Surf.Sci. {\em 345}, 85 (1996).\\

[17] D.Nattland, S.C.M\"uller, P.D.Poh and W.Freyland, J.Non-Cryst. Solids {\em 205--207},
772 (1996).\\

[18] W.B.Pearson, The Crystal Chemistry and Physics of Metals and Alloys, Wiley, New York, 1972.\\

[19] J.Als-Nielsen in: Structure and Dynamics of Surfaces, Topics in
Current Physics 43,
ed. by W.Schommers and P.von\,Blanckenhagen, Springer, Heidelberg, 1987.\\

[20] P.S.Pershan  and J.Als-Nielsen, Phys.Rev.Lett. {\em 52},759 (1984).\\

[21] M.J.Regan et al., Phys.Rev.B {\em 55},15874 (1997).\\

[22] P.Eisenberger, W.C.Marra, Phys.Rev.Lett. {\em 46}, 1081 (1981).\\

[23] H.Dosch, Critical Phenomena at Surfaces and Interfaces, Springer Tracts in
Modern Physics, Vol. 126, Springer, Heidelberg, 1987.\\

[24] M.J.Regan, H.Tostmann, P.S.Pershan, O.M.Magnussen, E.DiMasi, B.M.Ocko and
M.Deutsch,
Phys.Rev.B {\em 55}, 10786 (1997).\\

[25] S.A.Rice and M.Zhao, submitted (1998).\\

[26] T.B.Massalski et al. (eds.), Binary Alloy Phase Diagrams, ASM International, Materials Park,
Ohio, 1990.\\

[27] S.Dietrich, Wetting Phenomena, in: Phase Transitions and Critical Phenomena, Vol. 12, ed. by
C.Domb and J.Lebowitz, Academic Press, London, 1986. \\

[28] S.Dietrich, G.Findenegg and W.Freyland (eds.), Phase Transitions at Interfaces (Proceedings
Bunsen Discussion Meeting), Ber.Bunsenges.Phys.Chem. {\em 98} (1994).\\

[29] A.H.Narten, J.Chem.Phys. {\em 56}, 1185 (1972).\\

[30] H.Tostmann, E.DiMasi,O.G.Shpyrko, P.S.Pershan, B.M.Ocko, and M.Deutsch,
in preparation (1998).\\

[31] E.DiMasi, H.Tostmann, B.M.Ocko, P.S.Pershan and M.Deutsch,
in preparation (1998).\\

[32] S.Krimmel et al., Phys.Rev.Lett. {\em 78}, 3880 (1997).\\

[33] H.Reichert, P.Eng, H.Dosch and I.K.Robinson, Phys.Rev.Lett. {\em 74}, 2006
(1995).

\newpage
\subsection*{Figure Captions}

\noindent {\bf Fig.\,1} Sketch of the geometry of x-ray scattering
from the liquid surface with $\alpha$ and $\beta$ denoting incoming
and outgoing angle, the incoming and outgoing wavevector $k_{in}$
and $k_{out}$ respectively and the azimuthal angle $2\theta$. The
momentum transfer $q$ has an in-plane
component $q_{xy}$ and a surface-normal component $q_z$.\\

\noindent {\bf Fig.\,2} X-ray reflectivity from liquid Ga and liquid
Ga-In (16.5\% In) at 25$^{\circ}$\,C. The broken line represents the
fit of the model described in the text to the measured reflectivity
from liquid Ga. The solid line represents the fit of the same model
with one additional high density
adlayer to the measured reflectivity from the liquid  Ga-In eutectic.\\

\noindent {\bf Fig.\,3} X-ray reflectivity from Ga-Bi for three
different temperatures. The solid lines show the fit of the data to
a simple surface segregation model with one Bi monolayer on top of
the Ga rich bulk phase. The difference between the 35$^{\circ}$\,C
and the 95$^{\circ}$\,C data is due to the increased  roughness at
higher temperatures. The 260$^{\circ}$\,C data cannot be fitted by
this model. The inset shows the principal difference in surface
structure between the 35$^{\circ}$\,C and the 260$^{\circ}$\,C data,
indicating a wetting phase transition.

\noindent {\bf Fig.\,4} Comparison of the intrinsic or local real
space density profiles (thermal broadening removed) for Ga-Bi at two
different temperatures as well as for elemental Ga. The density is
normalized to the bulk density of liquid Ga. The zero position in
$z$ is arbitrarily assigned to the center of the first Ga layering
peak.

\noindent {\bf Fig.\,5} Grazing Incidence Diffraction from Ga-Bi at
35$^{\circ}$\,C. The solid line shows the in-plane structure
averaged over the penetration depth of x-rays away from the critical
angle for total external reflection (about 1000 {\AA}). The open
triangles show the diffraction peak obtained for $\alpha <
\alpha_{crit}$ ($\alpha_{crit}$ = 0.14$^{\circ}$ for Ga at $\lambda
= 0.653 {\AA}$) thus probing the surface region evanescently.

\noindent {\bf Fig.\,6} (a) X-ray reflectivity of Hg and
Hg/0.06at\%Au normalized to the Fresnel reflectivity of Hg. (b)
Normalized reflectivity of  a Hg/0.13\% alloy.

\newpage
\newpage
\begin{figure}[tbp]
\centering
\includegraphics[angle=90,width=1.0\columnwidth]{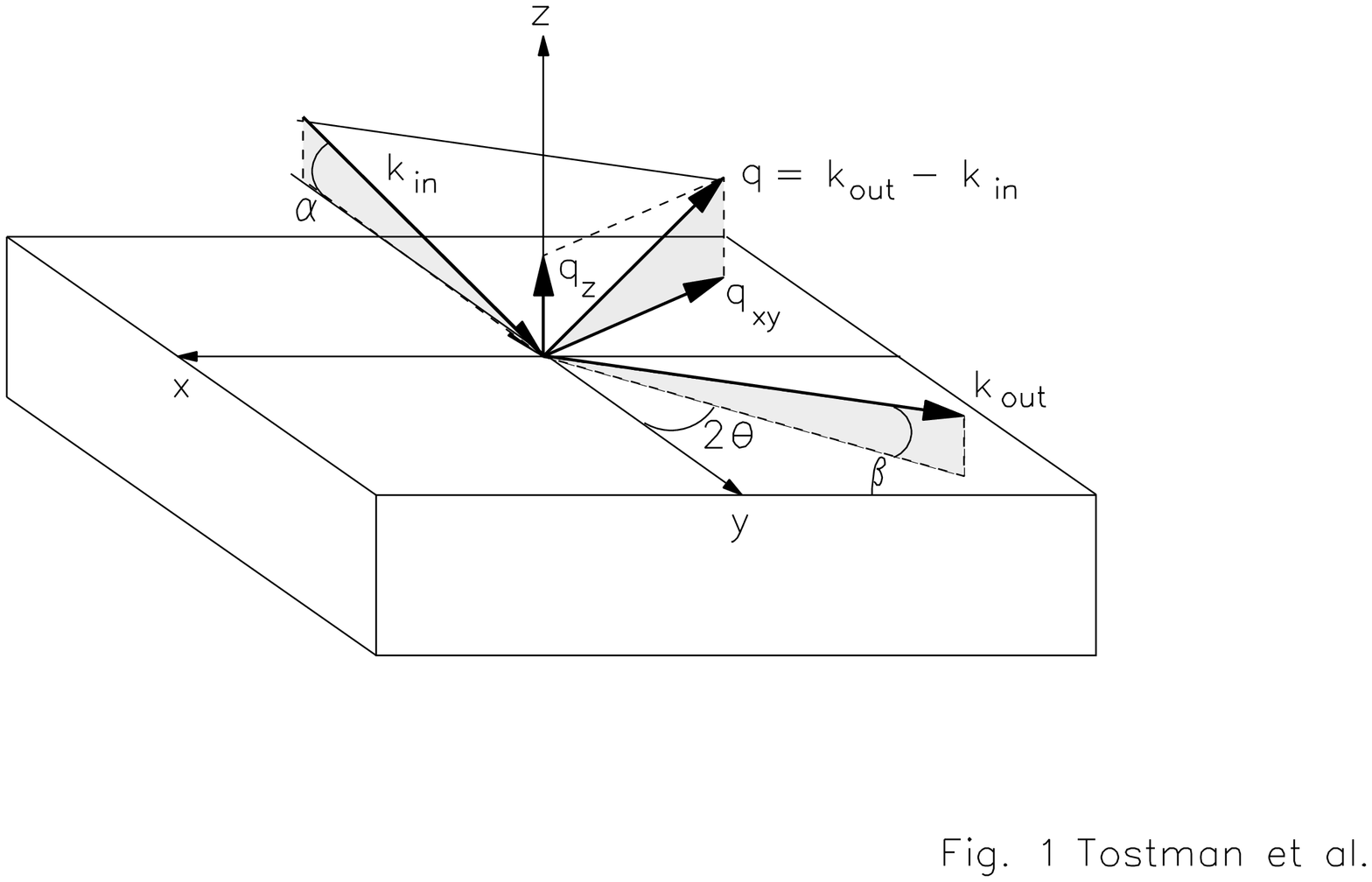}
\end{figure}

\newpage
\begin{figure}[tbp]
\includegraphics[angle=90,width=1.0\columnwidth]{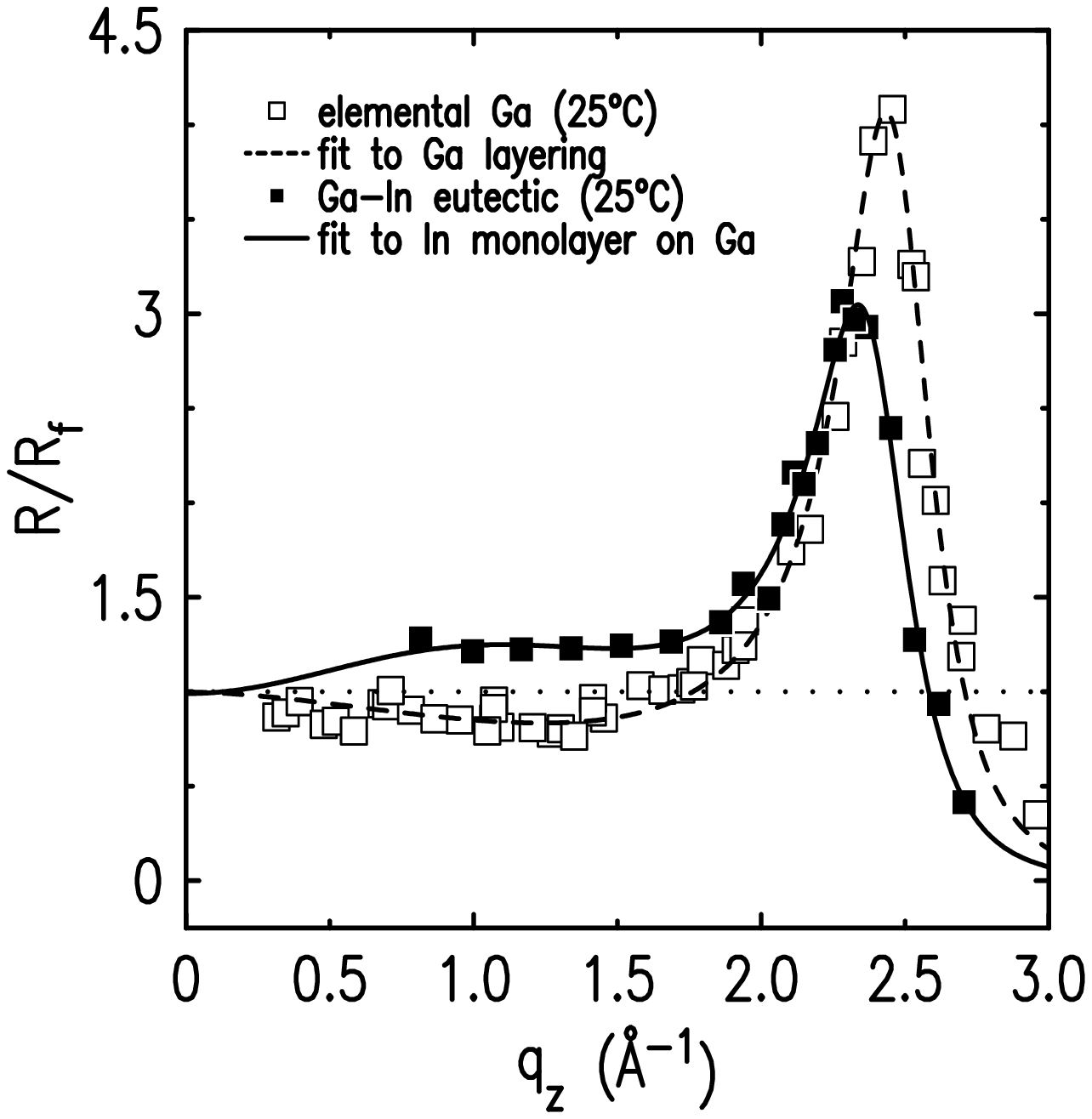}
\end{figure}

\newpage\newpage
\begin{figure}[tbp]
\centering
\includegraphics[angle=90,width=1.0\columnwidth]{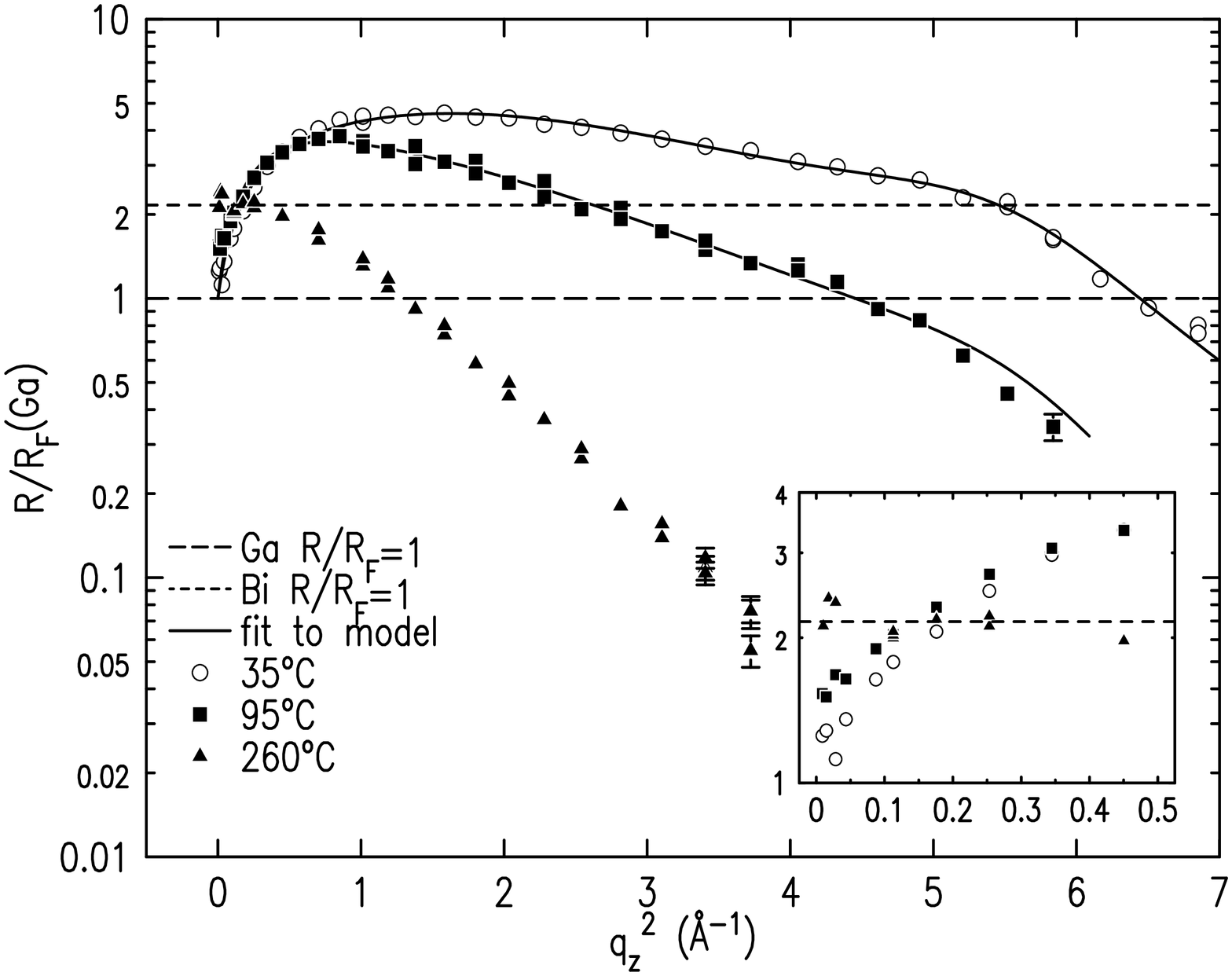}
\end{figure}

\newpage\newpage
\begin{figure}[tbp]
\centering
\includegraphics[angle=90,width=1.0\columnwidth]{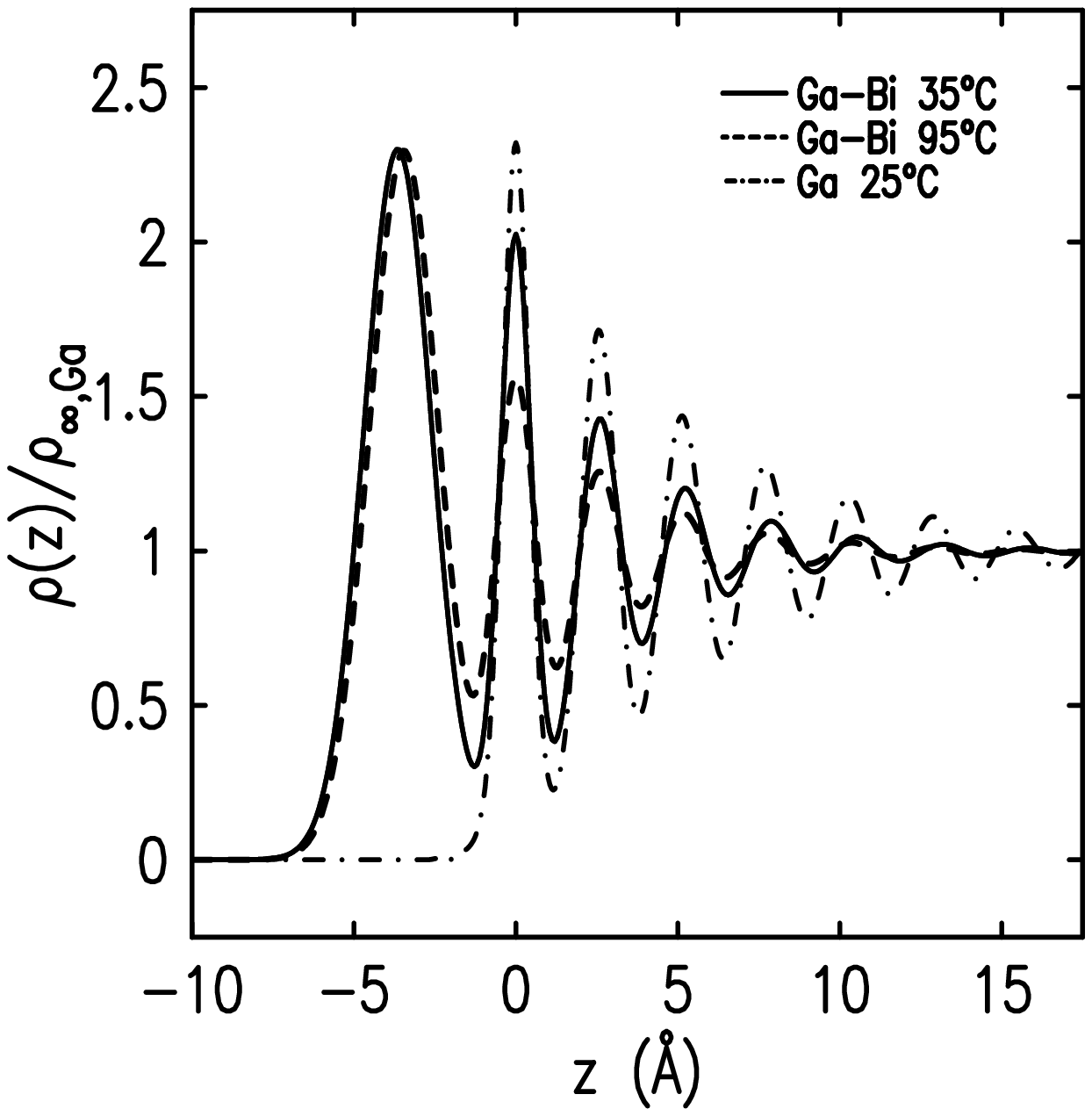}
\end{figure}

\newpage
\begin{figure}[tbp]
\centering
\includegraphics[angle=90,width=1.0\columnwidth]{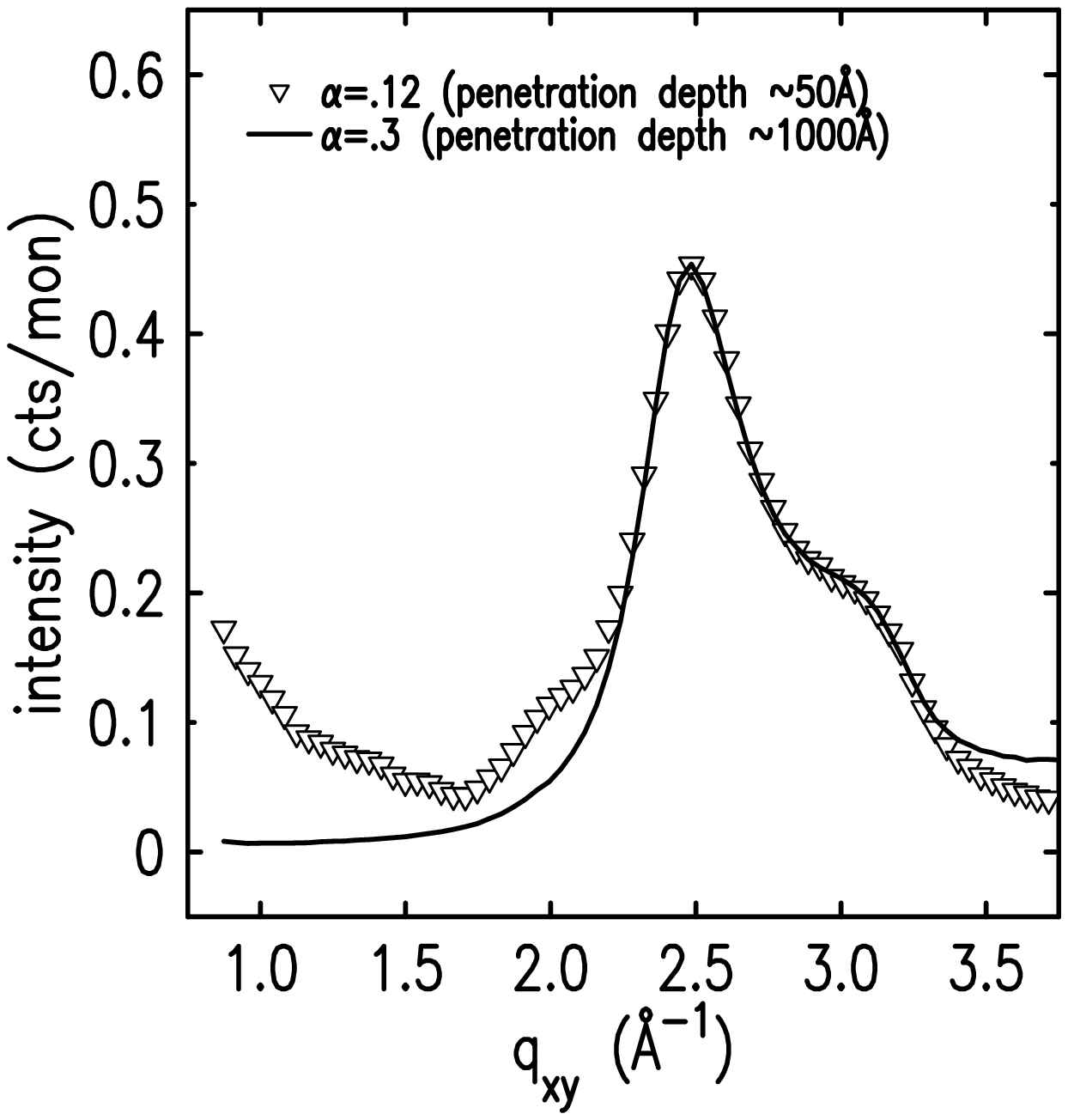}
\end{figure}

\newpage
\begin{figure}[tbp]
\centering
\includegraphics[angle=0,width=1.0\columnwidth]{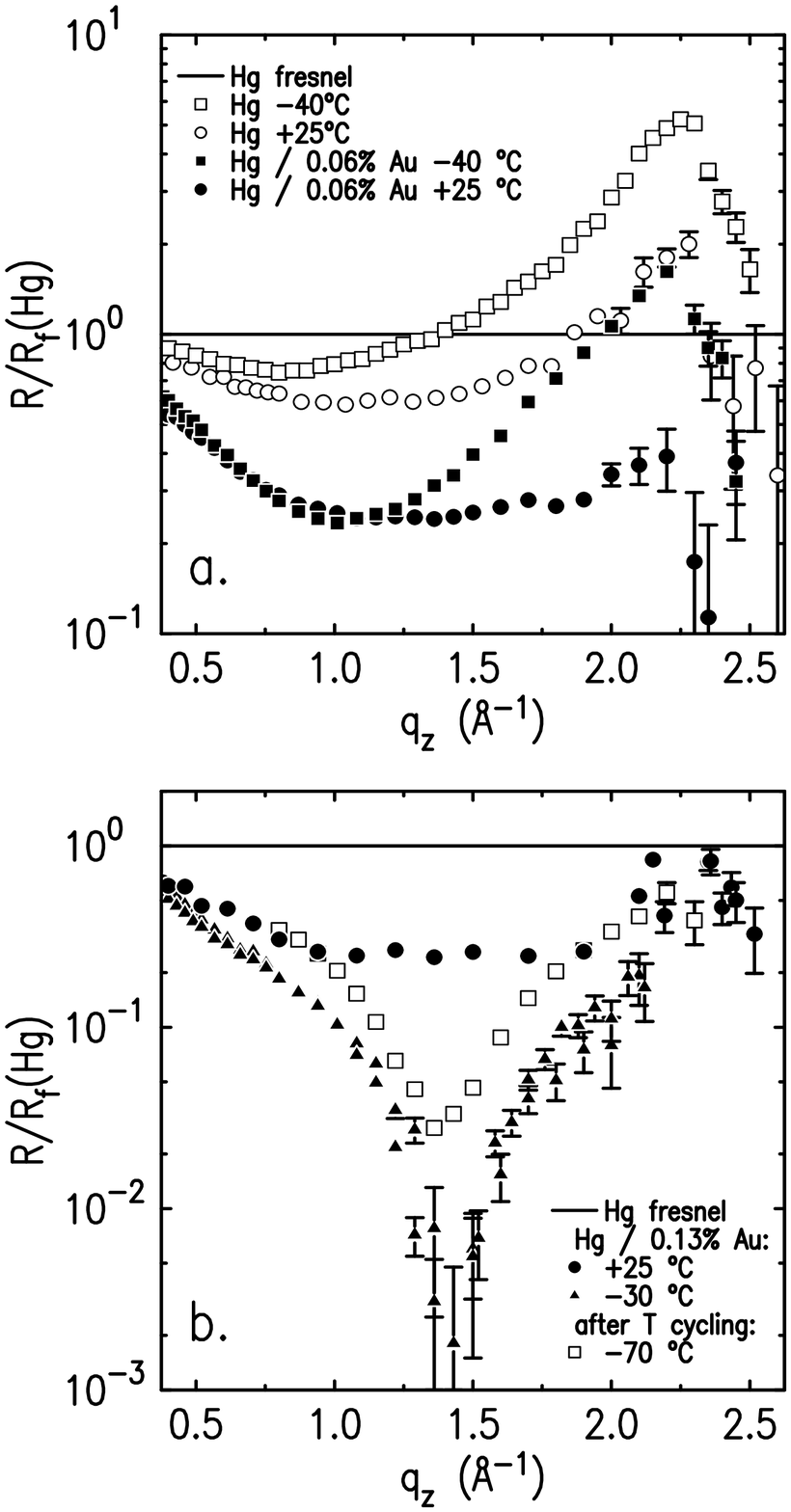}
\end{figure}

\end{document}